# Limitations of synthetic aperture laser optical feedback imaging

**Wilfried Glastre[*], Olivier Jacquin, Olivier Hugon, Hugues Guillet de Chatellus, and Eric Lacot**

*Centre National de la Recherche Scientifique / Université de Grenoble 1,*

*Laboratoire Interdisciplinaire de Physique, UMR 5588,*

*Grenoble, F- 38041, France*

*\*Corresponding author: wilfried.glastre@ujf-grenoble.fr*

In this paper we present the origin and the effect of amplitude and phase noise on Laser Optical Feedback Imaging (LOFI) associated with Synthetic Aperture (SA) imaging system. Amplitude noise corresponds to photon noise and acts as an additive noise, it can be reduced by increasing the global measurement time. Phase noise can be divided in three families: random, sinusoidal and drift phase noise; we show that it acts as a multiplicative noise. We explain how we can reduce it by making oversampling or multiple measurements depending on its type. This work can easily be extended to all SA systems (Radar, Laser or Terahertz), especially when raw holograms are acquired point by point.

## 1) Introduction

Making images with a good in-depth resolution through scattering media is a major issue, limited by a double problematic. Firstly the scattering medium generally strongly attenuates the ballistic photons signal which enables to obtain resolved images and the wavefront is highly perturbed by scattered photons, degrading the quality of the resolved image. Secondly concerning the accessible depth in samples, we are limited by the working distance of the objective. For the first issue, several ways to overcome these problems have been proposed among which we can distinguish two main families. The first one uses pre-compensation of the wavefront before propagation, to improve the resolution. This technique is successfully used both with optics or acoustic modalities [1,2,3], but it requires an *a priori* knowledge of the medium. The second one selects ballistic photons while rejecting multi-diffused parasitic photons: Optical Coherence Tomography (OCT) [4] and confocal microscopy associated [5] or not [6] to non linear effects belong to this family as well as tomographic diffractive microscopy [7]. Laser Optical Feedback Imaging (LOFI), based on optical reinjection in the laser cavity also belongs to this second family [8,9]. The principle of this technique is to use a laser both as a source and a detector of photons. By analyzing the coherent interaction between the emitted and reinjected photons, it is possible to know the complex amplitude and phase of the reinjected electric field. Amplitude [10] and/or phase [11] images can be obtained by scanning the object point by point with galvanometric mirrors or mechanical translations. We previously showed [12,13] that we benefit from an ultimate sensitivity at shot noise, that shows that LOFI is an excellent imaging system to makes images through scattering media, because of the weakness of the ballistic photons signal. In addition to that, LOFI has the advantage of being self-aligned and as a result, is very easy to implement. Concerning the solution to the second problem (accessible depth in samples), we shown in [14] that LOFI opens the way to another possibility: imaging beyond the objective working distance which is important to make deep images with high resolution. This is possible because LOFI gives both amplitude and phase information, therefore the blurred raw image from a scan beyond the working distance of the objective can be numerically refocused, keeping its initial numerical aperture. This is operation is called Synthetic Aperture (SA). This paper is a continuation of [14] where the case of an acquisition without any perturbation was presented. We now consider a more realistic case including noise during raw acquisition and we analyze its effects on the final synthetic images. Parasitic reflections occur on optical elements; we have shown in [15] that they can be divided in two groups: specular or diffusive and that, in absence of other noise, specular noise is constant and can be filtered out. As a result, diffusive parasitic reflections are the main limitation. In this paper, we are investigating the other kind of noises that could disrupt an acquisition and to simplify this study we suppose the absence of parasitic in what follows. More precisely, we first focus on laser quantum noise which is an additive noise. Then, we explore phase noise which can be divided in three families: random, sinusoidal and drift phase noises and which acts as a multiplicative noise. We identify their sources, assess their level and their consequences and propose several ways to handle them.

## 2) Reminder on our previous setup [14]

### Experimental setup

Our study is based on the LOFI experimental setup [14] and it is shown on Figure 1. Laser source system is highly sensitive to reinjected photons scattered by the target to be imaged. Informations about both amplitude and phase of reinjected electric field are accessible.

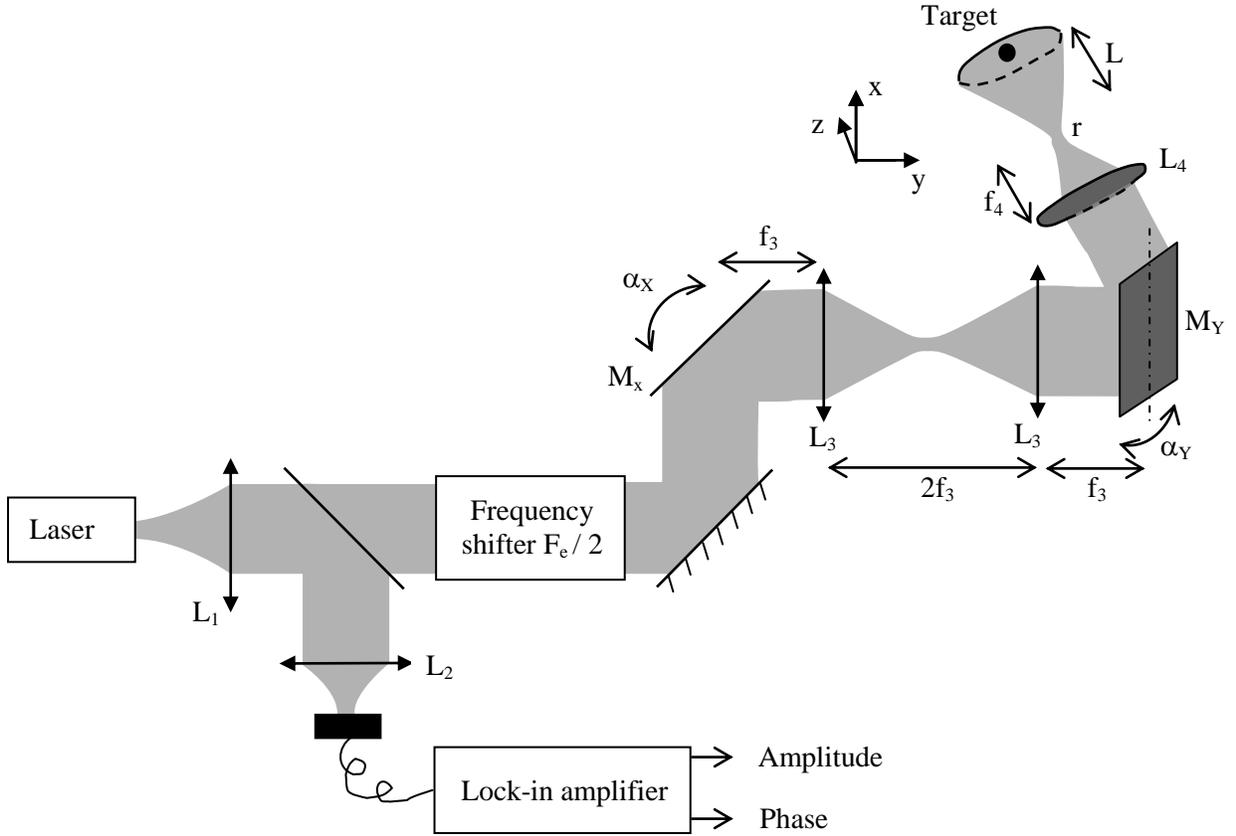

**Figure 1: Experimental setup of the synthetic aperture LOFI-based imaging system. The laser is a cw Nd:YVO$_4$ microchip collimated by lens L$_1$. A beam splitter sends 10% of the beam on a photodiode connected to a lock-in amplifier which gives access to the amplitude and phase of the signal. The frequency shifter is made of two acousto-optic modulators which diffract respectively in orders 1 and -1 and give a net frequency shift of F$_e$ / 2 = 1.5 MHz. X-Y plane is scanned by galvanometric mirrors M$_X$ (scan in the X direction) and M$_Y$ (scan in the Y direction) conjugated by a telescope made by two identical lenses L$_3$. f$_3$ and f$_4$ are the focal lengths of L$_3$ and L$_4$. α$_X$ and α$_Y$ are the angular positions of M$_X$ and M$_Y$. r is the waist of the laser after L$_4$.**

The image is obtained point by point by a 2D scanning of the two galvanometric mirrors M$_X$ and M$_Y$. The reason why we use galvanometric mirrors is to limit parasitic vibrations (leading to phase noise in the signal) and benefit from a quick movement compared to mechanical translational scanning. However, vibrations can not be totally eliminated and their consequences will be developed later in this paper. These mirrors are conjugated in the focal object plane of L$_4$ and as a result, when the mirrors are rotating, the beam scans the target with a translational movement. The scanning is made fast along one direction (X direction) and slowly along Y direction so the acquisition is made point by point and line by line.

The goal of our setup is to be able to get resolved images beyond classical working distance of the lens (or objective) $L_4$, that is why the target is placed at a distance L after the image focal plane of $L_4$ (Figure 1). Without any numerical treatment, by simply scanning the object in this configuration, we only get a raw complex defocused image. However we showed in [14] that using an appropriate numerical filtering, we are able to refocus this raw image into one with the resolution we would have if the object was in the image focal plane of $L_4$ (given by the beam waist r/√2). It is equivalent to say that we are able to artificially increase the working distance of $L_4$ while keeping its numerical aperture constant, at the price of a degradation of the photometric balance [14]. This numerical treatment is applied to raw image is Synthetic Aperture (SA) operation which is possible because we have both amplitude and phase information.

## Raw Point Spread Function (PSF)

By neglecting possible misalignment between the laser and the rotational axis of the two galvanometric mirrors, by using a defocus $L \gg Z_R = \dfrac{\pi r^2}{\lambda}$, the Rayleigh distance of the laser (far field condition) and by placing in paraxial condition (low aperture of $L_4$), we showed [14] that the raw acquisition of a punctual target is given by:

$$h_R(L, x, y) \propto \left( \exp\left(-\frac{x^2 + y^2}{2\,RES_R^{\,2}}\right) \exp\left( j\pi \frac{x^2 + y^2}{2\dfrac{L}{2}\lambda} \right) \right)^2 \quad (1)$$

$$RES_R = \frac{\lambda L}{\pi \sqrt{2}\, r}$$

This corresponds to a wavefront of lateral spatial width $RES_R(L)$ and a radius of curvature $L/2$. The Fourier transform of $h_R(L,x,y)$, is given by:

$$H_R(\nu,\mu) \propto \exp\left(-\frac{\nu^2+\mu^2}{\Delta\nu^2}\right)\exp\left(-j\frac{\pi L\lambda(\nu^2+\mu^2)}{2}\right)$$
$$\Delta\nu = \frac{\sqrt{2}}{\pi r}$$
(2)

In this expression (υ,μ) are the spatial frequencies associated with (x,y) and Δυ the spectral width of the raw signal.

### *Point Spread Function after Synthetic Aperture operation*

Because the raw signal corresponds to a wavefront defocused over a distance L / 2, we can recover the resolution by simply filtering the raw signal with $H_{filter}(L,\nu,\mu)$ the free space transfer function over a distance – L / 2. We shown is [14] that the final synthetic signal is given by:

$$|h_{SA}(x,y)| = |FT^{-1}(H_R(\nu,\mu)H_{filter}(\nu,\mu))| \propto \exp\left(-\frac{x^2+y^2}{RES_{SA}^2}\right)$$
$$RES_{SA} = \frac{r}{\sqrt{2}}$$
(3)

with $FT^{-1}$ the inverse Fourier Transform operation:

$$H_{filter}(\nu,\mu) = \exp\left(j\frac{\pi L\lambda(\nu^2+\mu^2)}{2}\right)$$
(4)

We finally recover a resolution $RES_{SA}$ ~ r whatever the initial defocus is (*i.e.* L). We are now going to study the effects on the final synthetic images of amplitude and phase noises. For the need of our demonstrations, we will use the object shown on Figure 2.

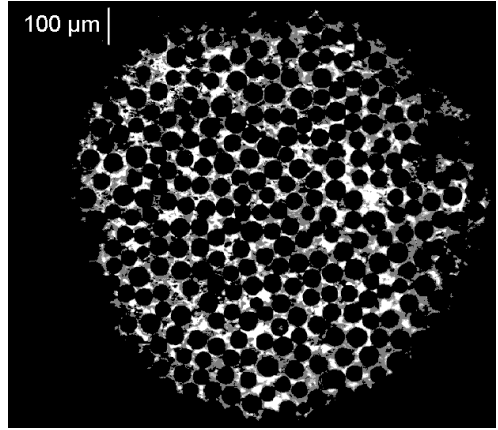

**Figure 2: Target used for the whole study: it is made of reflective silica beads of 40 μm diameter behind a circular aperture of 1 mm diameter. The bright field transmission image is made through a Zeiss microscope objective with a magnification of 10 and a 0.25 numerical aperture (focal length of 20 mm).**

## 3) Additive noise

Because of the LOFI sensitivity [12,13], this noise is mainly caused by the laser quantum noise and the detection is limited by the detection of one photon during the pixel integration time T.

*Problems and solutions to amplitude noise: theoretical analysis*

We are going to show that there are two main ways to reduce amplitude noise: an increase of the pixel integration time T or the spatial oversampling of the initial raw image (causing an increase of the number of pixels $N_{pixels}$). These two methods both increase the Signal to Noise Ratio (SNR) proportionally to the total acquisition time $T_{Total}$:

$$T_{Total} = N_{pixels} T \qquad (5)$$

We will now present the two methods.

### Increase of the integration time

Here we focus on the improvement we can get on the shot noise if we increase the integration time T while keeping the pixel number constant. The signal power (the square of the signal proportional to the flux of reinjected photons) does not depend on T whereas the noise power (proportional to the variance of the signal) is inversely proportional to T. As a result the signal to noise ratio in energy $SNR_{Integration}$ (ratio between the signal and the noise power) is proportional to T. Because $N_{pixels}$ is constant here, we get from Eq. (5):

$$SNR \propto T \propto T_{Total} \qquad (6)$$

### Oversampling of the raw image

We now focus on a second way to reduce the influence of the additive noise: the increase of the pixel number while the pixel integration time is kept constant. The random additive noise, in the Fourier field, spreads on the whole spectrum. This total spectrum is directly related to the sampling with the simple relations with $2\Delta\upsilon_{Shannon} = 1 / \delta x$ (size of total spatial spectral field recorded in directions X), with $\delta x$ the distance between two pixels in the X direction. Assuming that the sampling is the same in the X and Y directions, the surface of this Fourier spectrum is given by $S_{NoiseSpectrum} = 4\Delta\upsilon_{Shannon}^2 = 1 / \delta x^2$. However we can see from Eq. (2) and Figure 3, that the signal is localized over a surface (in the power spectral field) $S_{SignalSpectrum} = \pi\Delta\upsilon^2/2$ (factor 2 is because we are considering the Fourier power). As a result, it is possible to improve the final SNR by using an amplitude and phase filtering, instead of a pure phase filtering like in Eq. (4). By doing so, the major part of the signal information is preserved (only a factor 2 is lost corresponding to extreme plane waves) while most of the noise is rejected. This type of filter has already been used in SA-LOFI but in a rotational configuration [15,16]. If we want to optimize the SNR, the most appropriate filter is called the adapted filter (well known in Radar temporal field) and is given by H'$_{Filter}$($\upsilon$,$\mu$):

$$H'_{Filter}(\nu,\mu) = \exp\left(-\frac{\upsilon^2 + \mu^2}{\Delta\upsilon^2}\right)\exp\left(j\frac{\pi L\lambda(\nu^2+\mu^2)}{2}\right) \tag{7}$$

This filtering leads to the following synthetic signal:

$$\left|h'_{SA}(x,y)\right| = \left|TF^{-1}(H_R(x,y)H_{filter}(x,y))\right| \propto \exp\left(-\frac{x^2+y^2}{r^2}\right) \tag{8}$$

By comparing with Eq. (3), we can see that the interest of this filter for the photometric performances is obtained at the cost of a lower resolution of a factor $\sqrt{2}$. This can be explained by the fact that extreme plane waves in the signal are lost. More precisely, concerning the photometric performances, the use of this filter turns $S_{NoiseSpectrum}$ and $S_{SignalSpectrum}$ into $S'_{NoiseSpectrum} = \frac{\pi\Delta\upsilon^2}{2} = \frac{\pi\Delta\upsilon^2}{8\Delta\upsilon^2_{Shannon}} S_{NoiseSpectrum}$ and $S'_{SignalSpectrum} = S_{SignalSpectrum}/2$. We finally get an improvement in the SNR given by:

$$\frac{SNR_{AdaptedFilter}}{SNR_{PhaseFilter}} = \frac{\dfrac{S'_{SignalSpectrum}}{S'_{NoiseSpectrum}}}{\dfrac{S_{SignalSpectrum}}{S_{NoiseSpectrum}}} = \frac{S'_{SignalSpectrum}}{S_{SignalSpectrum}}\frac{S_{NoiseSpectrum}}{S'_{NoiseSpectrum}} = \frac{1}{2}\frac{8\Delta\upsilon^2_{Shannon}}{\pi\Delta\upsilon^2} = \frac{1}{\pi\Delta\upsilon^2\delta x^2} \propto N_{Pixels} \tag{9}$$

In this expression SNR$_{AdaptedFilter}$ and SNR$_{PhaseFilter}$ are respectively the SNR with or without adapted filter. Because of the constant integration time T for each pixel, the total measurement time is proportional to the spatial sampling and Eq. (9) can be written:

$$\frac{SNR_{AdaptedFilter}}{SNR_{PhaseFilter}} \propto N_{pixels} \propto T_{Total} \qquad (10)$$

Thus we have shown that whatever the method used to improve the SNR is, it is directly proportional to the total measurement time $T_{Total}$.

## *Experimental results*

We now illustrate the theoretical predictions with simulated and experimental data. We show on Figure 3 the Fourier transform amplitudes of a simulated PSF for different spatial samplings; this corresponds to Eq. (2).

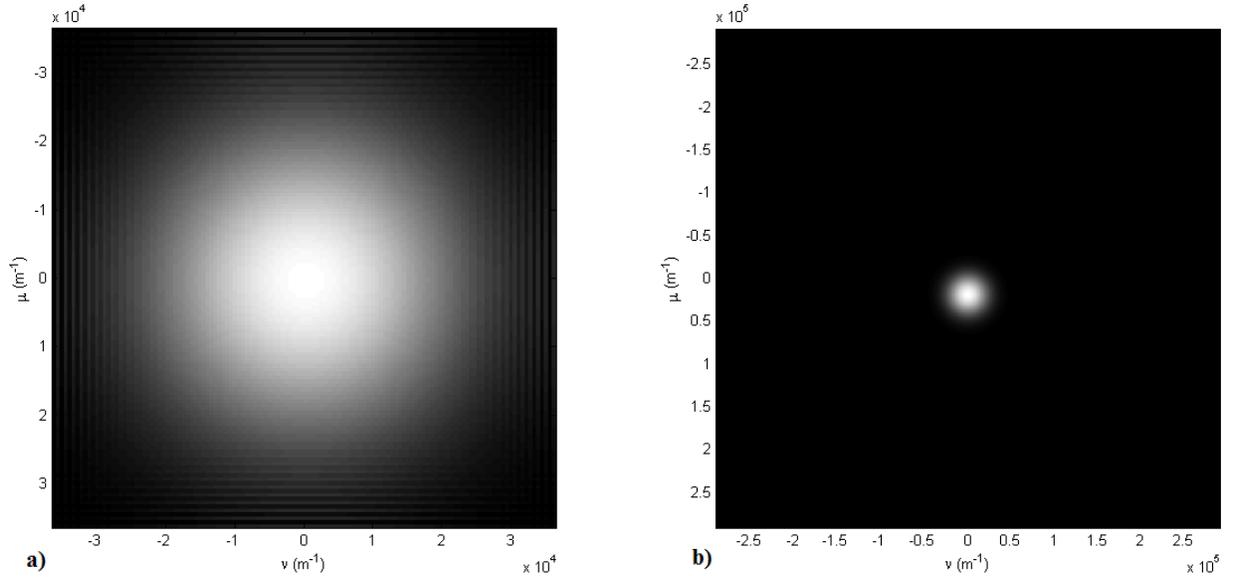

**Figure 3: Illustration of the effect of the spatial sampling on the Fourier content of the signal. The images are the amplitude of the Fourier transform of a simulated PSF with the following parameters: r = 20 μm, f = 75 mm and L = 2.5 cm. For a constant field image of 2 mm, we have a sampling of a) 128\*128 pixels and b) 1024\*1024 pixels.**

Figure 3 illustrates that the higher the sampling rate is, the stronger the signal isolation is in the total spectrum and consequently, the possibility to filter additive noise. We now show on Figure 4 the effect of the oversampling and of the use of adapted filtering on a real image of the object of Figure 2.

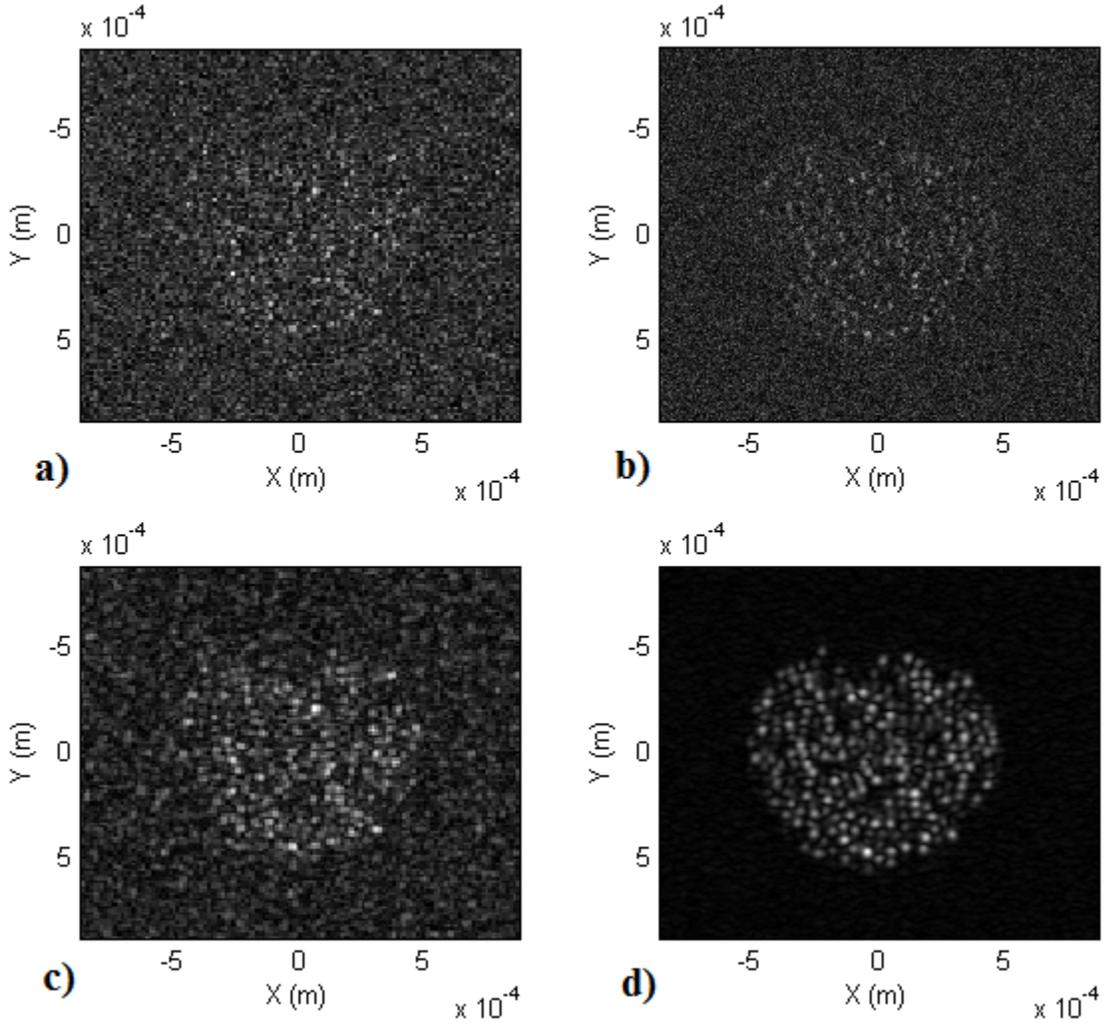

**Figure 4: Amplitude of SA images of the setup of Figure 2, parameters are r = 20 μm, f = 75 mm and L = 2.5 cm. Images a), c) show a sampling of 128*128 pixels and b), d) a sampling of 1024*1024 pixels. Figures are amplitude images after filtering a), b) by phase filter of Eq. (4) and c), d) adapted filter of Eq. (8).**

Figure 4 shows that the adapted filtering of an oversampled acquisition is a good way to improve the SNR. On Figure 5, we measure the evolution of the power (square of the amplitude normalized by the number of pixels) of both signal and noise when increasing the integration time (Figure 5a) or the pixel number combined with adapted filtering (Figure 5b).

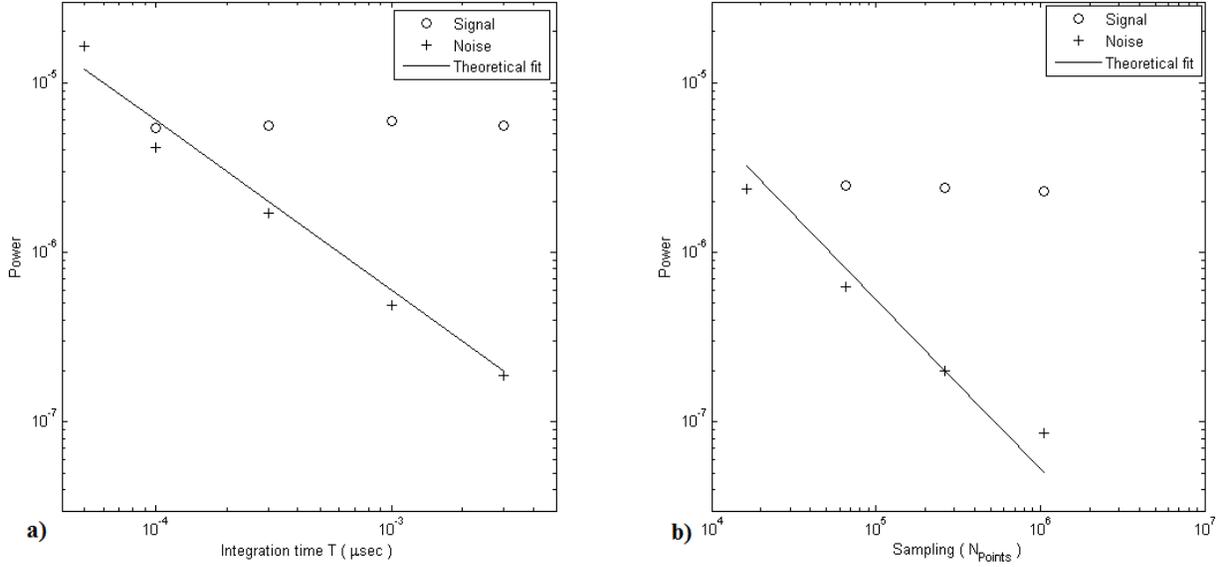

**Figure 5:** Dependence of the power in a pixel of signal and noise (averaged) in the SA image with the acquisition time. The signal is from the object of Figure 2 with parameters r = 20 μm, f = 75 mm and L = 2.3 cm. Acquisition time is increased via a) the integration time in a pixel at constant sampling and field of view or b) the sampling at constant integration time T and field of view. What we call power is the mean of the square of the image amplitude. This power is normalized by the total number of pixels. The noise is measured where there is no beads (see Figure 4).

Figure 5 illustrates the fact that when the total measurement time $T_{Total}$ is increased (by increasing T or $N_{pixels}$), the signal power remains unchanged while the noise is reduced proportionally to $T_{Total}$ which validates the theoretical predictions of the previous section. We can see that the signal power is divided by a factor 2 when adapted filtering is used (Figure 5b) which is conform to the theory. As a conclusion to this section, it remains preferable to increase the integration time instead of oversampling the signal, which slightly degrades the resolution.

## 4) Imperfections in phase acquisition (multiplicative noise)

In this section, another kind of noise that can affect a raw acquisition will be analyzed: the phase noise. Because it is a multiplicative noise, it impacts final synthetic images very differently: instead of being simply added to the ideal image, we will see that it turns a part of the signal power into parasitic noise which depends on the nature of the phase noise. In what follows, we will study the three main phase perturbations we have met: random phase noise, sinusoidal phase noise and phase drifts. Reminding that the raw acquisition of Eq. (1) is the expression of a wavefront of lateral spatial width $RES_R(L)$ and a radius of curvature $L / 2$, we are going to make analogies with well known physical situations to simplify our analysis and avoid tedious calculations.

### *Random phase noise*

This phase noise can have several origins: mechanical movements (ground, table, galvanometric mirrors), malposition of the fast mirror (X direction which correspond to the lines acquisition) between lines. In the case of the mechanical movements, this noise is independent for one pixel to another whereas in the case of the malposition, the phase noise will only be present between lines which will have a different effect on the final synthetic image. In our setup, mechanical noise produces a weak phase noise (lower than 0.01 radian, producing no visible effect) whereas malposition is stronger (about 0.1 radian).

#### **Theoretical analysis**

We represent the random phase noise by a random function $\Phi(x,y)$ with $P_\Phi(\phi)$ its associated density function. With this phase noise, the raw acquired wavefront $h_R(x,y)$ of Eq. (1) is turned into:

$$h_R^{PhaseNoise}(x, y) = m_{PhaseNoise}(x, y) h_R(x, y)$$
$$m_{PhaseNoise}(x, y) = \exp(j\ \Phi(x, y)) \tag{11}$$

In this expression $m_{PhaseNoise}(x,y)$ is the dephasing term. Physically, $h_R(x,y)$ corresponds to a wavefront generated by a waist $r / \sqrt{2}$ which have been propagated over a distance $L / 2$, SA filtering corresponding to refocusing back this signal. When we introduce phase defects $m_{PhaseNoise}(x,y)$ on the raw image, we simply generate speckle. More quantitatively, using a well known result about speckle, the mean square of our signal (our final synthetic image is random as the phase noise function is random) is given by [17]:

$$\overline{\left|h_{SA}^{PhaseNoise}(x,y)\right|^2} \approx \left|h_{SA}(x,y)\right|^2 * \left(\left|\overline{m_{PhaseNoise}(x,y)}\right|^2 \delta(x,y) + (\frac{2}{\lambda L})^2 DSP_m(\frac{2x}{\lambda L}, \frac{2y}{\lambda L})\right) \quad (12)$$

In this expression * is the convolution and $\overline{\phantom{xx}}$ the mathematical expectation operation. $DSP_m(\upsilon,\mu)$ is the power spectral density of $m_{PhaseNoise}(x,y) - \overline{m_{PhaseNoise}}$:

$$DSP_m(\upsilon,\mu) = FT(COV_m(x,y))$$

$$\overline{m_{PhaseNoise}} = \int_{-\infty}^{+\infty} \exp(j\phi) P_\Phi(\phi) d\phi = \tilde{P}_\Phi(1) \approx 1 - \frac{\sigma_\Phi^2}{2} \quad (13)$$

In this expression, $COV_m(x,y)$ is the covariance of $m(x,y)$ and $\tilde{P}_\Phi$ the characteristic function associated with random function $\Phi$ and $\sigma_\Phi$ its standard deviation. By analysing Eq. (12), we see that the SA operation divides the raw signal in two components: the first one is the signal we would have without any noise whereas the second one is the speckle term generated by the random phase noise on the raw signal. More precisely, the phase noise converts a part of the signal power into speckle, what is illustrated via the term $\overline{\left|m_{PhaseNoise}(x,y)\right|^2} = \left|\tilde{P}_\Phi(1)\right|^2 \approx \sigma_\Phi^2$ in Eq. (12) and (13); by conservation of the total energy from the raw signal, the proportion of the power in the speckle is therefore $1 - \left|\tilde{P}_\Phi(1)\right|^2 \approx 1 - \sigma_\Phi^2$. The greater the standard deviation of the random phase perturbation is, the higher the power conversion toward speckle is. More quantitatively, if we consider a Gaussian or a uniformly distributed phase noise, we get:

- For the Gaussian noise:

$$\overline{\left|m_{PhaseNoise}(x,y)\right|^2}\bigg|_{Gaussian} = \left|\tilde{P}_{\Phi,Gaussian}(1)\right|^2 = \exp(-\sigma_\Phi^2) \qquad (14)$$

- For the uniform noise:

$$\overline{\left|m_{PhaseNoise}(x,y)\right|^2}\bigg|_{Uniform} = \left|\tilde{P}_{\Phi,Uniform}(1)\right|^2 = \text{sinc}^2\left(\frac{\sqrt{12}\,\sigma_\Phi}{2}\right) \qquad (15)$$

Concerning spatial features of the speckle contribution, we see from Eq. (12) and (13) that it depends on the covariance of $m_{PhaseNoise}$: the narrower the covariance is, the wider the speckle pattern is as we can see on Figure 6. If the phase noise is independent from one pixel to another (case of mechanical noisy movements), the width of the covariance of $\Phi(x,y)$ is directly equal to the size of one pixel $\delta x$ and $\delta y$ in X and Y directions respectively. More quantitatively, the width of $DSP_{mm}(x,y)$ is $\sim 1/\delta x$ in X direction and so from Eq. (12) we deduce that the speckle pattern has a size $\sim \lambda L / \delta x$ in X direction (size of a beam diffracted over a distance L through a hole of size $\delta x$). As a result, at the minimum spatial sampling (Shannon limit: $\delta x \approx r$), the speckle has approximately the same size than the raw signal with a radius $RES_R(L)$.

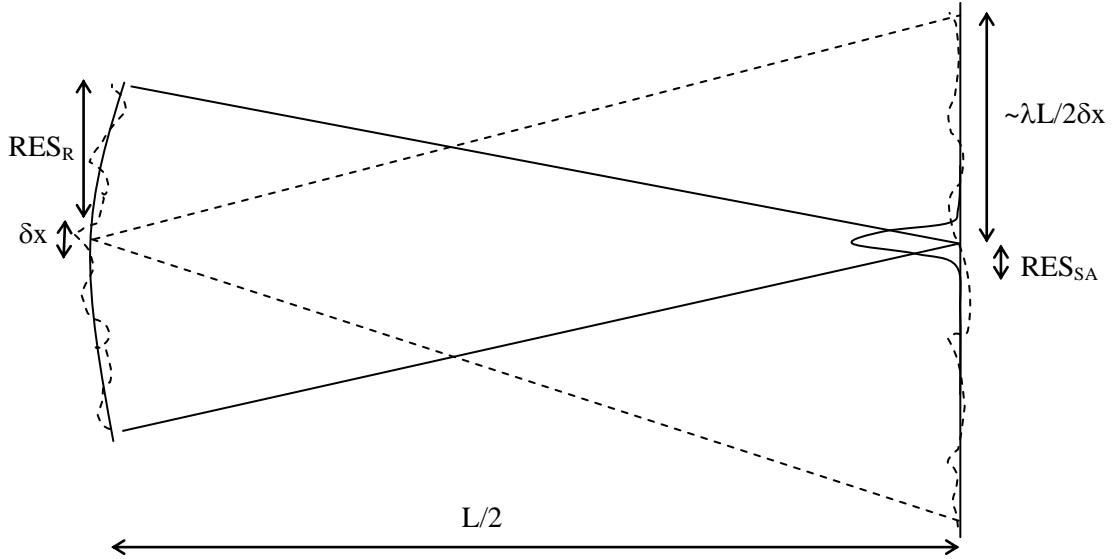

**Figure 6: Propagation of a wavefront with phase noise over a distance L/2. In the final image plane we have two contributions: a coherent one (plain line) and a random speckle (dashed line). The speckle and coherent contributions have relative intensities depending only on the density of probability of the random phase.**

In our case, as we said in the introduction of this section, the phase noise is mainly caused by a problem of malposition of the fast mirror which creates a phase noise only along the slow direction Y (there is a slight shift between lines). As a result the speckle is created only along this slow direction. Because we have estimated this noise around $\sigma_\Phi \approx 0.1$ radian, we expect from Eq. (14) that only 1% of the raw signal power is sent toward speckle while 99% of the power is kept for the synthetic final image. This good performance is the reason why we have chosen galvanometric mirrors to create a translational movement between the target and the laser instead of simply moving the object with a mechanical translational stage. Even if we have good performances, it is important to keep in mind that a phase noise of $2\pi$ (vibration amplitude of $\lambda/2$) is enough to totally convert our raw signal into speckle so phase noise remains a critical point that needs to be carefully handled.

As for the additive noise case, the SNR can be improved by filtering the speckle term of Eq. (12). Indeed, because it is spread in the whole Fourier space, the speckle can be reduced by spatial oversampling associated with an adapted filtering that will preserve the useful signal.

However, the power of the useful signal that has been converted into this speckle cannot be recovered.

## Numerical verifications

It is difficult to experimentally validate our theoretical predictions as we have shown that natural random vibrations are negligible. For those reasons, we have chosen to check the theory on simulated data. Figure 7 presents the effects of Gaussian random phase noise on SA final image:

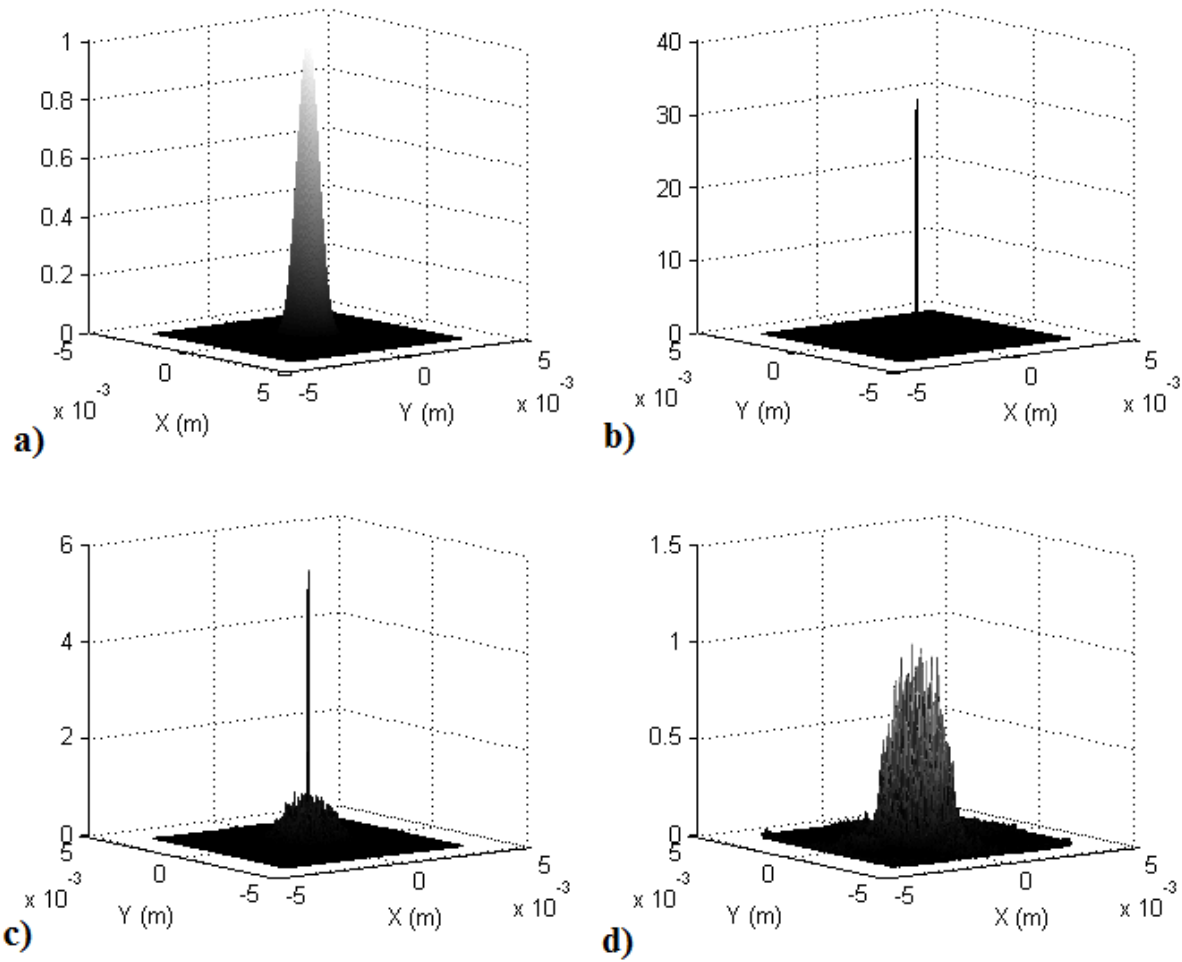

**Figure 7: Effect of random Gaussian phase noise on a SA operation. We use a simulated image of a punctual reflector. a) Amplitude of raw image with L = 4 cm, b) Amplitude after numerical refocusing, without phase noise, c) Amplitude after numerical refocusing, $\sigma_\Phi = 3\pi/5$ and d) Amplitude after numerical refocusing, $\sigma_\Phi = \pi$. Parameters are r = 20 μm, f = 75 mm and the definition is 512*512 pixels; the numerical refocusing is made with the pure phase filter for all images.**

We verify that, in accordance to Eq. (12) and (13), when introducing phase noise into raw acquisition, the power in the synthetic image is transferred into speckle noise (Figure 7b and Figure 7c). When phase noise $\sigma_\Phi$ exceeds $\pi$, the initial phase information is completely lost and all initial power in the raw image (Figure 7a) is turned into speckle (Figure 7d). We can see on Figure 7c and Figure 7d that the width of the speckle pattern is close to the width of the raw signal (Figure 7a) which is coherent with the theoretical considerations (Figure 6).

## *Sinusoidal phase perturbations*

We now focus on another important phase perturbation: sinusoidal phase noise. It has two main causes: the first one is the mechanical vibration of the table and of all optical components which is generally at a low frequency (< 300 Hz). The second source of sinusoidal noise is the electric power supply: 50/60 Hz and its harmonics that can be present and impact the galvanometric mirror motors. Globally these perturbations have an amplitude between 0 and 0.5 radian depending on the quality of our setup (measure of the phase evolution when galvanometric mirrors are at rest) and the attention we have paid to the source of vibration and to the electric shielding. In the same way as for the random phase noise, the repercussion of this perturbation on the final synthetic image is studied now.

### **Theoretical analysis**

Due to the scanning of the target, the sinusoidal temporal perturbation will correspond on the raw acquisition to a spatial sinusoidal perturbation. If we note $\Phi_0$ its amplitude and $(\upsilon_0, \mu_0)$ its spatial frequency, the raw signal is now given by:

$$h_R^{Sinusoidal\ Phase}(x,y) = m_{Sinusoidal\ Phase}(x,y) h_R(x,y)$$
$$m_{Sinusoidal\ Phase}(x,y) = \exp(j\Phi_0 \sin(2\pi(\upsilon_0 x + \mu_0 y)))$$
(16)

In this expression $m_{SinusoidalPhase}(x,y)$ is the perturbation term. Once again, to easily explain the effects of this term on the final synthetic image, it is more convenient to make a physical interpretation: adding the perturbation $m_{SinusoidalPhase}(x,y)$ is equivalent to insert a phase grating in front of the wavefront $h_R(x,y)$ before numerical refocusing (over a distance L /2). As a result, instead of having speckle, we now have several orders of diffraction and a repetition of several perfect synthetic images. Each of these images corresponds to an order of diffraction in our equivalent model of phase grating as illustrated on Figure 8. More precisely, the diffraction is along to the perturbation and the angles of diffraction are multiples of $\lambda\sqrt{\upsilon_0^2 + \mu_0^2}$ (see Figure 8). Because the SA filtering is equivalent to a retropropagation over a distance L / 2, the different orders are separated by a distance $\dfrac{\lambda L\sqrt{\upsilon_0^2 + \mu_0^2}}{2}$ on the final SA image:

$$h_{SA}^{Sinusoidal\ Phase}(x, y) = \sum_{n-\infty}^{+\infty} J_n(\Phi_0) h_{SA}(x - n\frac{\lambda \upsilon_0 L}{2}, y - n\frac{\lambda \mu_0 L}{2}) \qquad (17)$$

In this expression, $J_n(\Phi_0)$ is the Bessel function of order n. The proportion of the signal power sent in the order n is given by $|J_n(\Phi_0)|^2$ and we see that this expression is compatible with the total power conservation as $\sum_{n-\infty}^{+\infty}|J_n(\Phi_0)|^2 = 1$. As for random phase noise, the power of parasitic replicas (orders ≠ 0) is taken on the signal of interest (order 0).

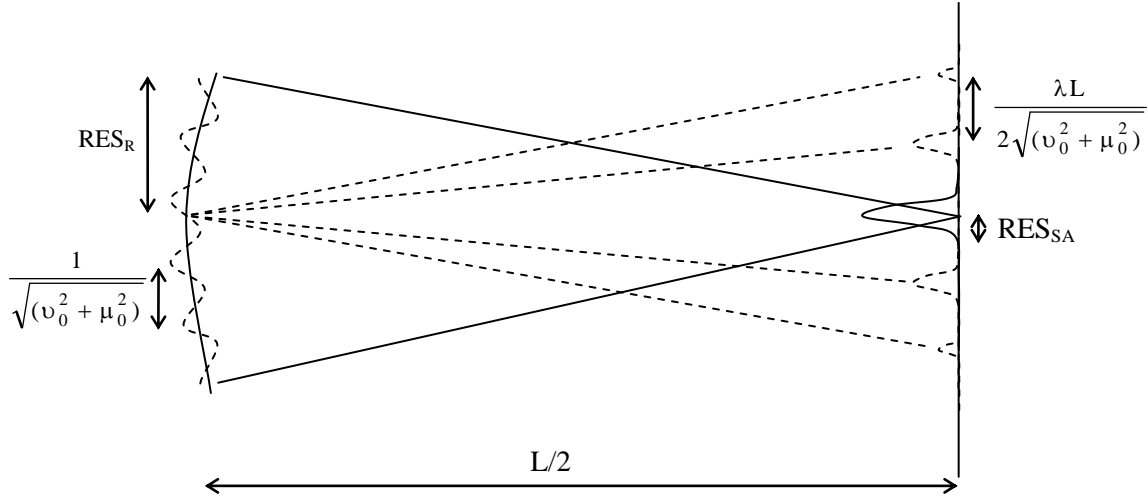

**Figure 8: Propagation of a wavefront with sinusoidal phase perturbations over a distance L/2. In the final image plane, there are two contributions: a coherent one (plain line) and several diffracted orders (dashed line). $\upsilon_0$ and $\mu_0$ are spatial frequencies of the perturbation in X and Y directions; the drawing is a projection along to the sinusoidal perturbation.**

In contrast with the previous perturbations (additive and random phase noises), oversampling and adapted filtering cannot reduce the image doubling effects.

Experimentally, we observe a vibration amplitude between 0 and 0.5 radian. According to Eq. (17), this corresponds to a transfer around 10% from order 0 (image we would get without noise) toward higher orders (parasitic replicas). As it was the case for random phase noise, an amplitude $\Phi_0$ around $\pi$ is enough to completely loose the phase information and the order 0 so it is very important to limit all sources of vibration and electric noise.

### Experimental check

To experimentally check the effects of the sinusoidal phase noise, we recorded a raw image (with a defocus of 2.5 cm) of the object of Figure 2 with or without imposing a mechanical vibration during acquisition. This vibration is imposed by an external loud speaker @ 100 Hz with an integration time T = 150 μs and a spatial sampling of 1.7 μm by pixel. Experimentally this creates a spatial frequency $\upsilon_0 \approx 10000$ m$^{-1}$ in the rapid direction (X). In the other direction (Y) we have measured $\mu_0 \approx 80000$ m$^{-1}$. The SA operation is then applied to recover the resolution. Synthetic amplitude images are presented on Figure 9:

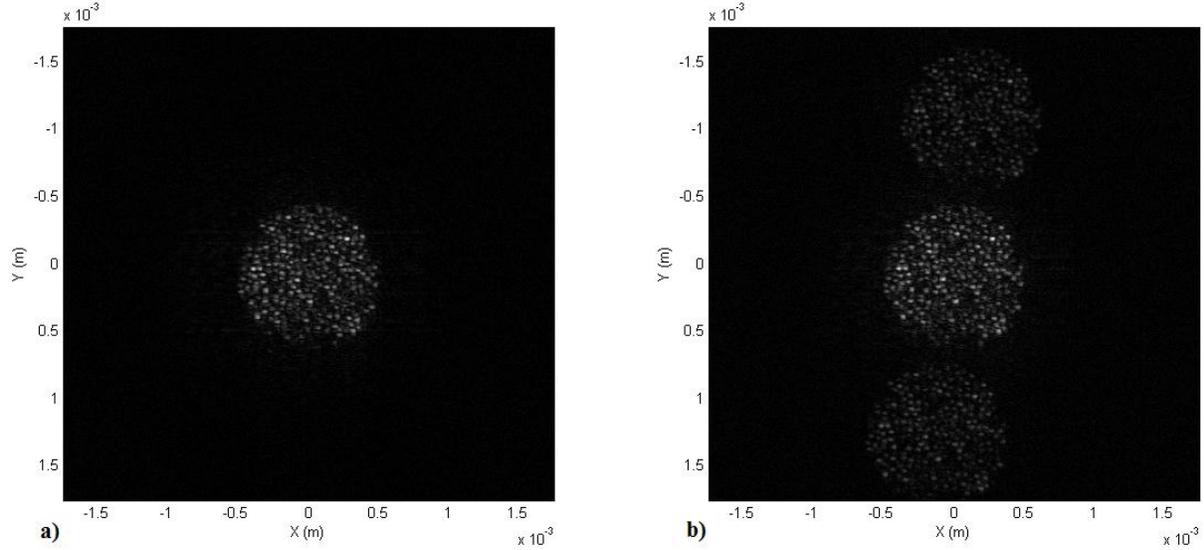

**Figure 9: Effect of mechanical sinusoidal phase perturbation on SA operation. Image parameters are 2048*2048 pixels, L = 2.5 cm, r = 20 µm, f = 75 mm, integration time T = 150 µs by pixel and the target is the object of Figure 2. Amplitude image after synthetic aperture operation a) without and b) with the perturbation. The perturbation at 100 Hz is generated by a loud speaker localized near the target. This induces a phase perturbation of amplitude $\Phi_0$ = 1.2 radian and of spatial frequencies $\upsilon_0$ = 10000 m$^{-1}$ and $\mu_0$ = 80000 m$^{-1}$. The SA operation is made with the pure phase filter of Eq. (4).**

Image replicas corresponding to diffraction orders can be observed on Figure 9 (here we see orders -1, 0 and 1). In theory, from Eq. (17) we expect a shift of $\lambda\upsilon_0 L/2$ = 130 µm and $\lambda\mu_0 L/2$ = 1.1 mm in X and Y directions respectively, what is conform to our experimental synthetic image on Figure 9b. Moreover, we have verified that the power distribution in different orders is given by Eq. (17).

### *Phase drifts*

There is a last possible phase perturbation: slow phase drift (compared to fast direction). This drift is mainly due to temperature fluctuations in the laser crystal when we turn on the laser or the fluctuations of the laser diode which is pumping. Another important cause is the variations of optical path between the laser and the target due to slight variation of the refractive index of the air (because of temperature or pressure changes). As a result, this phase drift can be highly reduced by isolating the system from atmospheric changes but it is difficult to completely eliminate it. In our case, the phase drift is around π radians/minute. We now present the impact of this phase drift on the final synthetic image.

### Theoretical analysis

Considering this perturbation, the raw signal can now be written:

$$h_R^{PhaseDrift}(x,y) = m_{PhaseDrift}(y) h_R(x,y)$$
$$m_{PhaseDrift}(y) = \exp(j\,\Phi(y))$$
(18)

We can see that the phase perturbation depends only on the Y coordinate which is the slow direction. This can be explained by the fact that we consider the case of a slow phase drift. As for the two previous phase perturbations, we can make a physical analogy: instead of a ground glass (random phase) or a grating (sinusoidal noise) in front of equivalent wavefront, the slow phase drift is introducing optical aberrations. As a result, the final synthetic image will only be distorted along the Y direction, depending on the precise aberrations which have been introduced. More quantitatively, as it was the case for the two previous phase noises, one part of the power in the center of the synthetic PSF (Eq. (3)) is lost proportionally to $\sigma_\Phi^2$ (variance of the aberration) and transferred into adjacent pixels which enlarge this synthetic image. Thanks to this ascertainment, we see that a drift of $2\pi$ is enough to highly degrade the final synthetic image. In the light of this information, because our acquisition time is around the minute and reminding that the phase drift is around π radians/minute, we see that this phase perturbation is very critical for us and that we need to correct it. We propose an efficient solution to do that: instead of making only one acquisition with a quick scan along the X direction leading to Eq. (18) and which is imperfect along the Y direction, we make a second acquisition but with a quick scan along Y this time. We then have two images: one without drift along X and the other along Y, but by combining them, we can finally recover a corrected "raw" acquisition before applying SA filtering and getting an aberration-free synthetic image.

## Experimental verification

To experimentally illustrate our theoretical considerations, we acquired two raw images (one with a quick acquisition along the X and the other along the Y direction) of the object of Figure 2 with a defocus L = 2 cm with a high phase drift. The phase drift is accentuated by an external perturbation of the laser diode (which pumps the laser crystal) power supply in order to have a phase variation of more than $\pi$ over a length $RES_{SA}$ (that is to have significant impact on the final synthetic image, see Eq. (3)). What we obtain is shown in Figure 10:

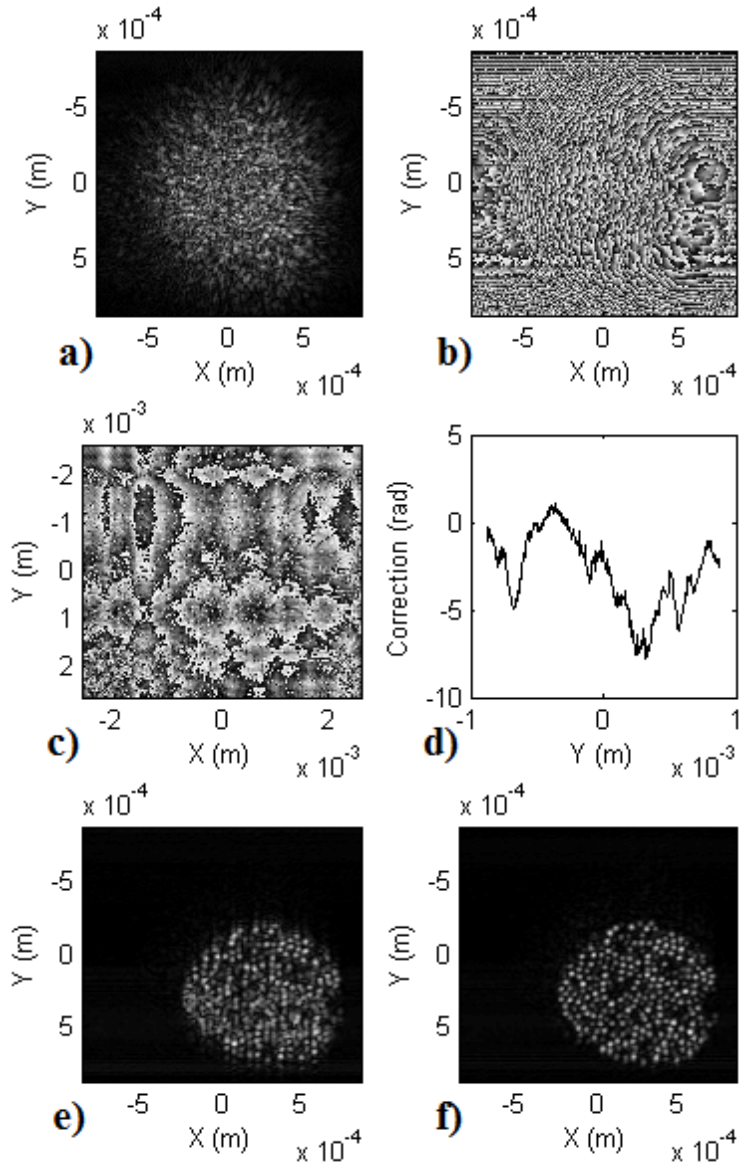

**Figure 10: Effect of phase drifts during the raw acquisition on SA imaging. Parameters are r = 13 μm, f = 25 mm, L = 2 cm and 512*512 pixels. The effect of our phase drift correction is illustrated too (here the correction is made on the image taken slowly along Y direction). The target is still the object of Figure 2. a) Amplitude and b) phase (white is -π radians and the black is +π) of raw image of the target. Image c) shows phase difference between the two raw acquisition acquired with different slow directions. d) is the phase correction to apply in the Y direction calculated from c). The two last images are amplitudes of the synthetic image (pure phase filter is used) e) before and f) after phase correction.**

On Figure 10e we can see the amplitude of synthetic image from one of the two raw acquisitions (precisely when Y is the slow direction, Figure 10a is its amplitude and Figure 10b its phase). We verify that, in accordance to the theory, the phase drift causes vertical aberrations. Figure 10c is the phase difference between the two images before any correction; we see there is a phase drift in both directions because of the two different slow directions used during acquisition. By using this image, we can correct the first raw image (with drift along Y) by eliminating vertical phase difference between the two images. This phase correction to apply is, of course independent with the column and so Figure 10c is averaged along X (Figure 10d). Finally, when SA operation is applied to the corrected image, we get Figure 10f which is free of aberration, showing that our correction method is efficient.

## 5) Conclusion and perspectives

In this paper, we have continued the work that was presented in [14], we presented a Synthetic Aperture LOFI-based setup which aims to make image deep through scattering media. Here we have explored the main sources of noise that can impact the acquisition, their repercussion on final synthetic images and proposed solutions to limit their influence. More precisely we have divided noises into two families: the additive (amplitude) noise and the multiplicative (phase) noise. The first is due to shot noise and can be reduced (relatively to the power SNR) proportionally to the global time measurement by increasing the integration time by pixel T or by oversampling image during acquisition and use adapted filtering. The second can still be divided into three sub-families: random phase noise mainly caused by galvanometric mirror malposition from one line to another, sinusoidal phase noise caused by mechanical vibrations and phase drifts caused by slow variations of temperature and pressure in the setup. Because they are multiplicative noises, they all convert a power fraction $1 - \sigma_\Phi^2$, where $\sigma_\Phi$ the mean noisy phase excursions, of the signal (useful signal and parasitic reflections) into parasitic signal which depends on the precise nature of the perturbation. This noise is pretty low but it is important to keep in mind that $\sigma_\Phi \sim \pi$ can be sufficient to completely destroy the phase information. That is why phase noise can be catastrophic if not controlled. Concerning random phase noise, this parasitic signal can be compared to speckle and can be partially reduced by oversampling and adapted filtering (as for amplitude noise). Sinusoidal phase noise is like introducing a grating which splits the useful signal in several orders each order corresponds to an image replica. Finally, phase drifts lead to aberrations in the direction of slow acquisition which can be corrected by combining two images with different "slow directions". The study we made in this paper is related to our previous work [14,15] but can easily be generalized to all interferometric imaging systems and especially those with a raw acquisition which is made point by point. This is the case for most of other SA systems whatever is their wavelength range : Radar [18,19], optical [20,21] or more recently Terahertz [22].

To put it in a nutshell, beside the signal there is a lot of noise sources that need to be limited: specular (can be filtered because it is constant) and diffusive parasitic reflections [15], shot noise, and noise converted from signal (useful and parasitic reflections) by phase noise. Because our goal is to make images through scattering media, our main challenge is to realize images with minimum number of photons. In this case, noise converted from useful signal by phase noise can be neglected but we see that close to the ultimate limit (shot noise), noises related to parasitic reflection are still present (diffusive or/and specular associated to the phase noises). A solution was proposed in [23], consisting in tagging photons with an acoustic transducer just in front of the target, thus eliminating parasitic reflections from the signal. Unfortunately, the proposed setup gives only access to the amplitude of the reinjected signal. Our future work aims to adapt it in order to recover the phase which is needed for Synthetic Aperture operations.